# An Order Statistics Post-Mortem on LIGO–Virgo GWTC-2 Events Analyzed with Nested Sampling

*Talya Klinger\* and Michalis Agathos*

**The data analysis carried out by the LIGO–Virgo collaboration on gravitational-wave events utilizes nested sampling to compute Bayesian evidences and posterior distributions for inferring the source properties of compact binaries. With poor sampling from the constrained prior, nested sampling algorithms may misbehave and fail to sample the posterior distribution faithfully. Fowlie et al. (2020) outlines a method of validating the performance of nested sampling, or identifying pathologies such as plateaus in the parameter space, using likelihood insertion order statistics. Here, this method is applied to nested sampling analyses of all events in the first and second gravitational wave transient catalogs (GWTC-1 and GWTC-2) of the LIGO–Virgo collaboration. The insertion order statistics are tested for uniformity across 45 events in the catalog and it is found that, with a few exceptions that have negligible effect on the final posteriors, the data from the analysis of events in the catalog is consistent with unbiased prior sampling. There is, however, weak evidence against uniformity at the catalog-level meta-test, yielding a Kolmogorov–Smirnov meta-$p$-value of $1.44 \times 10^{-3}$.**

## 1. Introduction

Since the first direct detection of gravitational waves (GWs) in 2015,[1] the LIGO–Virgo collaboration (LVC) has published the detection of tens of GW signals emitted by coalescing black-hole and neutron-star binaries, in the three observing runs carried out so far (O1, O2, and O3).[2–4] In gravitational-wave data analysis, parameter estimation is the process of inferring the source properties of signals which have already been identified as gravitational waves produced by compact binaries. Bayesian inference methods are employed to fit waveform models to data, using algorithms designed for efficiently sampling high-dimensional parameter spaces. One such algorithm is nested sampling, a method for efficiently computing Bayesian evidences as well as posterior probability distributions, introduced by Skilling in 2006[5] (for a review, see ref. [6]). Here, we use a new method of statistically verifying nested sampling output, the insertion order cross-check developed by Fowlie et al.,[7] to test for biased nested sampling in the LVC's gravitational-wave data analysis.

The LVC uses nested sampling alongside Markov Chain Monte Carlo (MCMC) and occasionally RIFT[8] to obtain posterior distributions in parameter estimation. These algorithms serve necessary and complementary purposes. While the LVC's implementation of MCMC converges faster for signals with long inspiral times, and RIFT allows for direct comparison to numerical relativity, they do not directly compute evidence, delivering only the normalized posterior and requiring further statistical calculations to estimate the evidence, introducing significant statistical errors. Nested sampling computes the evidence directly, allowing for greater accuracy.

In the LSC algorithm library (LAL), nested sampling was originally implemented in the *LALInference* package.[9,10] During the third observing run (O3), the LVC (now LVK) Collaboration gradually shifted its main data analysis pipelines to a newer Bayesian inference library *bilby*,[11] a Python-based modular code which combines LAL's libraries for data infrastructure and waveform modeling with third-party nested samplers. The insertion order cross-check defined later in this paper has already been implemented in many *bilby* samplers, including *CPNest*,[12] *nessai*,[13] and *UltraNest*.[14] However, the first two observing runs, O1 and O2, and the first half of the third observing run, O3a, were analyzed using *LALInference* alone. In this work, we utilize the insertion order cross-check to perform a post-mortem analysis on all nested sampling output for GW events in O1, O2, and O3a, evaluating the validity of parameter estimation results for the LVC's event catalogs GWTC-1 and GWTC-2.[15]

This article is structured as follows. In Section 2 we briefly describe the nested sampling algorithm and the insertion order statistics we use to estimate the validity of the LVC analyses. We then describe our implementation for the parameter-estimation

T. Klinger
Cardiff University School of Physics and Astronomy
5 The Parade, Newport Road, Cardiff CF24 3AA, UK
E-mail: talyaklinger@gmail.com

M. Agathos
DAMTP
Centre for Mathematical Sciences
University of Cambridge
Wilberforce Road, Cambridge CB3 0WA, UK

M. Agathos
Kavli Institute for Cosmology Cambridge
Madingley Road, Cambridge CB3 0HA, UK

**DOI:** 10.1002/andp.202200271

dataset output by *LALInference* in Section 4 and present the results on the GW events in Section 5. Concluding remarks are given in Section 6.

## 2. Insertion Order Statistics in Nested Sampling

### 2.1. Nested Sampling

For a given GW event associated with the coalescence of a compact binary, we can describe its source properties by a parameter vector $\theta \in \Theta$, where $\Theta$ denotes the corresponding parameter space, including the mass and spin of each component, the distance to the source, its sky-location and orientation angles, time, and phase of coalescence, as well as any additional parameters relating to matter properties in case of a neutron star, orbital eccentricity, etc. Given the observed data $D$, our aim is to infer the parameters $\theta$ of the source, i.e. estimate the posterior distribution $P(\theta|D, \mathcal{I})$ under the assumption that our background information $\mathcal{I}$ about the nature of the source, the behavior of our detectors and the validity of GR as the underlying theory is correct.

In Bayesian statistics, this amounts to updating our prior expectations quantified by $P(\theta|\mathcal{I})$ by making appropriate use of Bayes' theorem

$$P(D|\theta, \mathcal{I}) \times P(\theta|\mathcal{I}) = P(D|\mathcal{I}) \times P(\theta|D, \mathcal{I}) \quad (1)$$
$$L(\theta) \times \pi(\theta) \, d\theta = Z \times p(\theta) \, d\theta$$

$L(\theta) = P(D|\theta, \mathcal{I})$, known as the likelihood function, and $\pi(\theta) = P(\theta|\mathcal{I})$, the prior, give the desired quantities $Z = P(D|\mathcal{I})$, the evidence and $p = P(\theta|D, \mathcal{I})$, the posterior. Computing the likelihood function (the probability density for observing data $D$, given the model and the true values of the parameters) requires models for both the detector signal and noise—in LIGO's case, *LALSimulation* can generate a waveform model for the signal, while the noise for each detector is assumed to be Gaussian and is characterized by a power spectral density (PSD) which is pre-estimated based on a stretch of data around the time of the event.[15] Information from all detectors in operation is combined into a coherent network likelihood which is the product of individual detector likelihoods.[10] The task of efficiently sampling the parameter space to map the likelihood function, is carried out by the nested sampling algorithm.

The evidence $Z$—the probability of observing the measured data, given the model—is defined as

$$Z = \int L(\theta)\pi(\theta) d\theta \quad (2)$$

This is an important quantity in Bayesian data analysis, as the evidences produced by different models can be directly compared. Hence, the evidence can be used to rank competing hypotheses and quantify how much a given model is supported by the data. $dX = \pi(\theta)d\theta$ is known as the element of prior mass. If the prior mass contained by a likelihood contour

$$X(\lambda) = \int_{L(\theta)>\lambda} \pi(\theta) d\theta \quad (3)$$

is known, the evidence can be written as a 1D integral,

$$Z = \int_0^1 L(X) dX \quad (4)$$

which is more computationally manageable than integrating across a high-dimensional parameter space $\Theta$.

Nested sampling is a method for computing evidence that takes advantage of this formulation, relying on the statistical properties of prior sampling to provide a fast and accurate estimate of the prior mass at each integration step.

### 2.2. Summary of the Nested Sampling Algorithm

Nested sampling relies on sampling from the constrained prior: points from the prior with likelihood higher than some minimum value. As points from the constrained prior are sampled and discarded throughout the algorithm, the samples used at each step are called live points.

The nested sampling algorithm proceeds as follows:

1. Choose the number of live points $n_{\text{live}}$ and sample $n_{\text{live}}$ initial points from the constrained prior. Also, set an evidence threshold $\epsilon$.
2. Identify the live point with the lowest likelihood $L_i^*$. Discard the live point and record its likelihood.
3. Sample a new live point from $\pi(\theta)$ with $L > L_i^*$. At this stage, the prior volume compresses exponentially, giving prior volume $X_i \approx \exp(-1/n_{\text{live}})$ on the $i$th step (the proof is nontrivial, see ref. [5]).
4. Integrate the evidence $Z_i$ using $L_i^*$ and $X_i$.
5. Repeat steps (2)–(4) until a stopping condition is reached: $L_{\max} X_i / Z_i < e^\epsilon$, where $L_{\max}$ is the highest likelihood discovered so far, $X_i$ is the prior volume inside the current iso-likelihood contour $L_i^*$, and $Z_i$ is the current estimate of the evidence. For *LALInference*, $\epsilon = 0.1$; essentially, if all the live points were to have the maximum discovered likelihood, the evidence would only change by a factor of less than 0.1.[10]

Nested sampling requires faithful sampling from the constrained prior to produce accurate evidences and posteriors. In practice, sampling from the entire prior and accepting only points with high enough likelihood is impractically slow, because the volume of acceptable points decreases exponentially in time. So, most implementations of nested sampling sample from a restricted region of parameter space drawn around the live points. *LALInference*, in particular, generates samples by an MCMC chain from a randomly chosen previous livepoint, and choosing the length of the MCMC chain is a tradeoff between speed and accuracy.[10]

If the restricted region is too small or the MCMC chains too short, the constrained prior may not fully cover the iso-likelihood contour, violating the fundamental assumptions of nested sampling. Plateaus—regions of constant $L(\theta)$—also violate the assumptions of nested sampling, causing live points to be nonuniformly distributed in $X$.

### 2.3. Insertion Order Crosscheck

The insertion index is the position where an element must be inserted in a sorted list to preserve order. More concretely, if $x$ is a sorted list and there exists a sample $y$ such that

$$x_{i-1} < y < x_i \quad (5)$$

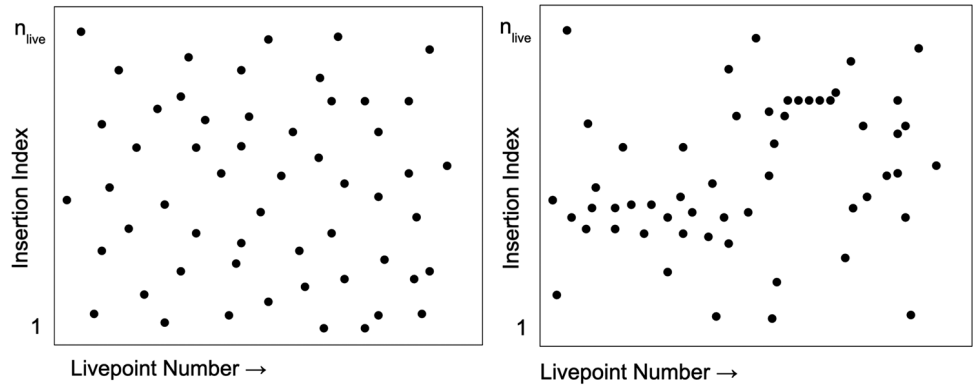

**Figure 1.** These illustrations depict likelihood insertion order plots from well-behaved (left) and pathological (right) nested sampling runs. In the first plot, the points, each representing a likelihood insertion order, are evenly distributed throughout the plane. In the second plot, some regions are densely populated with points or almost empty, associated with excessive sampling from a specific likelihood range. There is also a run of repeated indices, resulting from a plateau in prior space.

the insertion index of $y$ in list $x$ is $i$.[7] For example, in the list (1 3 5 7 9), $y = 4$ would have an insertion index of 3.

Fowlie et al.[7] noted that, if the assumptions of nested sampling are met, the insertion index of new live points into the list of likelihoods of current live points should follow a uniform distribution; that is, new live points should have an arbitrary likelihood, only constrained to be higher than the lowest likelihood. The prior mass enclosed by a certain likelihood decreases monotonically as that likelihood increases, so sorting live points by likelihood is equivalent to sorting by prior mass.

Therefore, nonuniformity of likelihood insertion indices serves as an early warning for any irregularities in sampling the prior. For example, a likelihood plateau in the parameter space results in a stretch of repeated indices. Such a problem would be visible in plots of the likelihood insertion index or detectable with statistical tests of uniformity. **Figure 1** illustrates what such nonuniformities might look like in a plot of insertion order.

For our purposes, the insertion order cross-check is the simplest and most flexible method for verifying nested sampling. Buchner's "shrinkage test" is limited to toy likelihood functions with certain analytic properties, and is designed for constrained prior sampling using regions (such as ellipsoids) rather than *LALInferenceNest*'s MCMC steps.[16] The diagnostic methods implemented in *Nestcheck*[17] are more applicable to LIGO parameter estimation, but they require multiple runs. One of the strengths of the insertion order crosscheck is that it can provide useful information on any scale, from an individual chain to an entire observing run.

## 3. Gravitational Wave Data

During the first three observing runs, two LIGO detectors (LIGO Hanford, LIGO Livingston) and the Virgo detector participated in the network. During O1, only the two LIGO detectors were operating, with Virgo beginning operation in August of 2017 during O2. In this work, we analyze 45 events in total, which were observed by two or more detectors as listed in **Table 1** (with the exception of one single-detector event GW190424A). In all three observing runs, the data used is sampled at 4096 Hz for all detectors, with a low frequency between 20 and 30 Hz, and upper frequency determined by the Nyquist frequency. The noise PSDs are calculated using either off-source data (*LALInference*) or on-source data (*BayesWave*) methods, while several calibration and data-cleaning methods were employed; these techniques are described in detail in refs. [15, 18, 19]. To account for possible miscalibration in the instruments, calibration errors were marginalized over, following the uncertainty envelopes in amplitude and phase provided by the data quality team.

### 3.1. First Observing Run (O1)

In the first observing run, O1, extending from 12 September 2015 to 19 January 2016, three GW events were detected with high confidence, all of which were identified as signals from the coalescence of binary black holes (BBH): GW150914 (the first gravitational-wave detection), GW151012, and GW151226. For the catalog paper,[18] all O1 events were analyzed using two waveform models: IMRPhenomPv2,[20] a phenomenological model calibrated to numerical relativity for gravitational waves from precessing BBH binaries, and SEOBNRv4, a model based on the effective one-body formalism.[21] We have analyzed the nested sampling results of all three events in O1.

### 3.2. Second Observing Run (O2)

O2 includes seven more BBH mergers and one binary neutron star (BNS) merger, all observed between 30 November 2016 and 25 August 2017. Some significant milestones from O2 are the first BNS event, GW170817, and the first GW signal to be detected by both LIGO interferometers and the Virgo interferometer, GW170814. The O2 events were also analyzed using IMRPhenomPv2 and SEOBNRv4, with a few exceptions. GW170729 was also analyzed with IMRPhenomD (a model similar to IMRPhenomPv2 for spinning but nonprecessing binaries), and GW170809 was analyzed using IMRPhenomD and multiple runs of IMRPhenomPv2 with different priors. The events from O1 and O2 together formed the first GW transient catalog GWTC-1, the results of which are detailed in ref. [19]. Our data crosscheck includes all available output files for the primary nested sampling runs of the ten binary black hole mergers. We omit the binary neutron star merger since it was not analyzed with nested sampling.

**Table 1.** Event-level meta-*p* values for each event in chronological order, together with the set of interferometric detectors (IFOs) that participated in the detection, the number of chain files per event, and number of live points used in the analysis.

| Event | IFOs | #chains | $N_{live}$ | Meta *p*-value |
|---|---|---|---|---|
| GW150914 | HL | 12 | 2048 | 0.2953 |
| GW151012 | HL | 12 | 1024 | 0.0973 |
| GW151226 | HL | 12 | 2048 | 0.0973 |
| GW170104 | HL | 12 | 1024 | 0.1047 |
| GW170608 | HL | 12 | 1024 | 0.0333 |
| GW170729 | HLV | 20 | 2048 | 0.2519 |
| GW170809 | HLV | 32 | 2048 | 0.0212 |
| GW170814 | HLV | 26 | 2048 | 0.1603 |
| GW170818 | HLV | 16 | 1024 | 0.1419 |
| GW170823 | HL | 18 | 2048 | 0.0317 |
| GW190408A | HLV | 4 | 2048 | 0.5709 |
| GW190412A | HLV | 4 | 2048 | 0.8341 |
| GW190413A | HLV | 4 | 2048 | 0.4058 |
| GW190413A | HLV | 4 | 2048 | 0.0387 |
| GW190421A | HL | 8 | 2048 | 0.2037 |
| GW190424A | L | 4 | 2048 | 0.3008 |
| GW190503A | HLV | 4 | 2048 | 0.1350 |
| GW190512A | HLV | 4 | 2048 | 0.2218 |
| GW190513A | HLV | 16 | 2048 | 0.4650 |
| GW190514A | HL | 4 | 2048 | 0.2093 |
| GW190517A | HLV | 4 | 2048 | 0.5249 |
| GW190519A | HLV | 4 | 2048 | 0.6515 |
| GW190521A | HL | 4 | 2048 | 0.3208 |
| GW190521B | HLV | 4 | 2048 | 0.6330 |
| GW190527A | HL | 4 | 2048 | 0.0330 |
| GW190602A | HLV | 4 | 2048 | 0.9461 |
| GW190620A | LV | 4 | 2048 | 0.0321 |
| GW190630A | LV | 4 | 2048 | 0.5345 |
| GW190701A | HLV | 4 | 2048 | 0.0849 |
| GW190706A | HLV | 7 | 2048 | 0.1111 |
| GW190707A | HL | 4 | 2048 | 0.3056 |
| GW190708A | LV | 4 | 2048 | 0.1964 |
| GW190719A | HL | 4 | 2048 | 0.1297 |
| GW190720A | HLV | 4 | 2048 | 0.1349 |
| GW190727A | HLV | 4 | 2048 | 0.3458 |
| GW190728A | HLV | 4 | 2048 | 0.7932 |
| GW190731A | HL | 4 | 2048 | 0.4246 |
| GW190803A | HLV | 4 | 2048 | 0.8563 |
| GW190828A | HLV | 5 | 2048 | 0.1731 |
| GW190828B | HLV | 4 | 2048 | 0.5201 |
| GW190909A | HL | 4 | 2048 | 0.1586 |
| GW190910A | LV | 4 | 2048 | 0.4375 |
| GW190915A | HLV | 4 | 2048 | 0.6632 |
| GW190929A | HLV | 4 | 2048 | 0.6984 |
| GW190930A | HL | 4 | 2048 | 0.1731 |

### 3.3. First Half of Third Observing Run (O3a)

The next upgrade to the instrumentation of both LIGO detectors and Virgo further improved their sensitivity, increasing the distance reach and thus the event detection rate of the three-detector network. In O3a, the first half of the third observing run between 1 April 2019 and 1 October 2019, 39 new GW events were detected with high confidence, of which we include 35 events analyzed with *LALInferenceNest* in this analysis. Among the most interesting events in O3a are a couple of highly asymmetric (in mass) compact binaries, the first black-hole—neutron-star binary candidates and the most massive black-hole binary observed to date, reaching a total mass of $\approx 150 M_\odot$. Several different waveform models were employed in the analyses of these events, especially for the ones whose mass ratio was found to be highly asymmetric, as well as some that showed weak signs of spin-induced precession. Both phenomenological (IMRPhenom) and effective-one-body (SEOBNR) models were used in all cases; however, here we focus on the nested sampling analyses that used IMRPhenomPv2 as the underlying waveform model, a set-up that is used in the analyses of almost all O3a events. Further details about the data, the detection statistics and the properties of all O3a events can be found in ref. [15]. The events from O3a, together with the ones from O1 and O2, form the second gravitational wave transient catalog, GWTC-2.[22,23]

## 4. Implementation

### 4.1. Data Parsing

The likelihood insertion indices defined in Section 2.3 can be computed either from the nested sampling iteration *i* and likelihood, or birth and death contours (initial and final likelihood), of each point. *LALInference* stores both the contours and the likelihoods themselves, but this information is distributed across two different types of output files. Log files contain the nested sampling iteration *i*, birth and death contours, and likelihood of each point. Since the likelihood is stored to higher precision than the birth and death contours, we use the likelihood and iteration to compute insertion indices. However, the initial pool of $N_{live}$ (typically 1024 or 2048) points is missing from the log files, so it is impossible to compute the insertion indices of early live points from the logs alone.

On the other hand, the main data product of *LALInferenceNest*, the chain files, contain the initial live points but not the nested sampling iteration, making it impossible to compute insertion order from these files alone. Whenever possible, we match each log file with the associated chain file to find the initial live points, then replay the algorithmic process to compute the insertion indices exactly. In the absence of a match, we attempt to minimize the effect of the missing initial $N_{live}$ live points by removing the first $5 \times N_{live}$ points from our reconstructed chain (a cutoff which removed missing-point effects from insertion order plots effectively), then compute the insertion indices for the remaining points.

Log files have less strict I/O specifications than chain files and occasionally have some duplicate blocks of points, resulting from resubmitted jobs. When this is the case, we identify and remove the older duplicate live points.

### 4.2. Measuring Uniformity

To interpret the computed insertion order statistics, we must measure how uniformly they are distributed between 1 and $N_{live}$. Several statistical tests exist to determine whether two (or more) samples are drawn from the same underlying probability distribution, or compare a sample to a reference probability distribution. In particular, we use the Kolmogorov–Smirnov (KS) test,[24] as implemented in *scipy*.[25]

The KS test measures the distance between two cumulative distribution functions (CDFs). More precisely, for empirical CDF $F_{data}(x)$ and CDF of the uniform distribution $F_U$, the Kolmogorov–Smirnov statistic is defined as

$$D_n = \sup_x |F_{data}(x) - F_U(x)| \qquad (6)$$

The KS test produces a test statistic between 0 and 1, with higher values corresponding to more distinct distributions. The KS statistic is independent of the number of samples, so the number of samples must be taken into account separately when interpreting the test results.

In the context of hypothesis testing, the *p*-value of a measurement or test statistic $x$ is the probability of obtaining the observed value, or a more extreme value (for uniformity testing, a test statistic associated with a larger difference between the two distributions), assuming that the null hypothesis is correct. The KS test can be converted to *p*-values through the Kolmogorov–Smirnov distribution, which associates probabilities with test results for the hypothesis that two samples are drawn from different distributions. In cases where nested sampling has proceeded correctly, the insertion order distribution is uniform and the null hypothesis is true, leading to a uniform distribution of *p*-values over different runs. In isolation, small *p*-values do not necessarily mean that an entire nested sampling run is compromised, but if small *p*-values predominate, that could be a sign of systematic problems.

A seemingly insignificant subtlety that turns out to be important is that the standard KS test implementation is designed for comparing distributions of continuous variables; however, in this case our insertion order data is discrete.[14] The impact of this effect is discussed in Section 5. There are two ways of resolving the problem: we can either transform the data into an equivalent continuous distribution ranging in $[1, N_{live} + 1)$ by adding a random number in $[0, 1)$ to each insertion index (which will respect uniformity if the underlying discrete distribution is uniform); or we can implement a discrete version of the KS test or variations thereof, as described in refs. [26, 27]. For simplicity, we have chosen to do the former.

### 4.3. Performing Insertion Order Crosscheck

*LALinference* is heavily parallelized, with each event's nested sampling analysis split into several parallel chains. The top and bottom panels of **Figure 2** showcase two examples of the insertion order distribution of single chains, one very uniform, one less so.

For each event, we compute insertion orders and KS statistics over each individual chain. To assess the overall quality of nested sampling in each event, we perform a KS test on the combined insertion orders from all parallel chains which we will refer to as the "event-level meta test".

In each chain, we also perform "rolling tests": a series of tests on each sequential stretch of $2 \times N_{live}$ points. An example rolling test is shown in Figure 4. These rolling test values can be used to examine and compare the severity of local anomalies. In particular, we report the minimum *p*-value from all rolling tests of each event. To assess the uniformity of the entire GWTC-1 and GWTC-2 dataset, we perform a final KS test on the KS *p*-values from each individual rolling test, known as the "catalog-level meta test."

## 5. Results

We perform the nested sampling replay on each chain and reconstruct the insertion index data for each event in Table 1. We first perform rolling KS tests for each chain and verify that none of the individual chains shows significant bias in its insertion order statistics. For each one of the events we then pool together the insertion indices from all parallel chains and perform a single event-level KS test by calculating the KS-statistic for that event and the corresponding *p*-value. The *p*-values for all the events analyzed are given in Table 1. We find a seemingly healthy distribution that spans the entire range between 0 and 1, whose uniformity we will examine with a higher level meta-*p*-value test for the entire catalog of events.

Before describing the final catalog-level results, let us first discuss the impact of two subtle effects that we introduced in Section 4: the effect of the discrete nature of the insertion index and its "continuification;" the effect of missing data and our recovery process.

Although the insertion index is a discrete variable, it is plausible to assume that its wide range ($N_{live}$ is typically 2048) effectively makes it continuous in practice. However, due to the very large number of data points, from the perspective of a KS test the underlying discreteness is significant. The KS statistic tests uniformity by measuring the maximum deviation of the empirical CDF from the diagonal. The empirical CDF of any discrete variable will inevitably have a staircase-like structure, and the ability of the KS test to discern this deviation from uniformity depends on the range of the discrete variable ($N_{live} \approx O(10^3)$) and the sample size. For large sample sizes (here $O(10^5)$), the discreteness has a significant impact, as our results demonstrate. When treating the insertion index as a continuous variable, without transforming it to one, we found a clear deficit at high *p*-values, with no event in the range [0.8,1.0] and a very high excess in the low *p*-value range [0,0.2]. Accounting for the discrete nature of the variable remedies this bias to a large extent.

Next, we examine whether our attempt of correcting for missing data, either in the initial $N_{live}$ points or in mid-run, has a significant effect on the overall insertion order statistics. On a per-chain basis, the removal of $5 \times N_{live}$ samples on either side of the gap seems to suppress the bias completely (even for a small number of missing chain points, the chain-level *p*-value is typically corrected by several orders of magnitude, back to O(0.1)). However, the catalog-wide statistics point in a different direction: in the population of all chains in the catalog, a residual bias remains. The size of this effect can be estimated by completely removing all chains that had missing data from the catalog-level analysis. Although these make up a small fraction of the total number of chains, we find that their removal leads to a systematic improvement of the meta-$p$-value by a factor of ≈4. A possible compromise would be to increase the number of points we truncate around missing data, however this in practice would render the heavily truncated chains virtually uninformative.

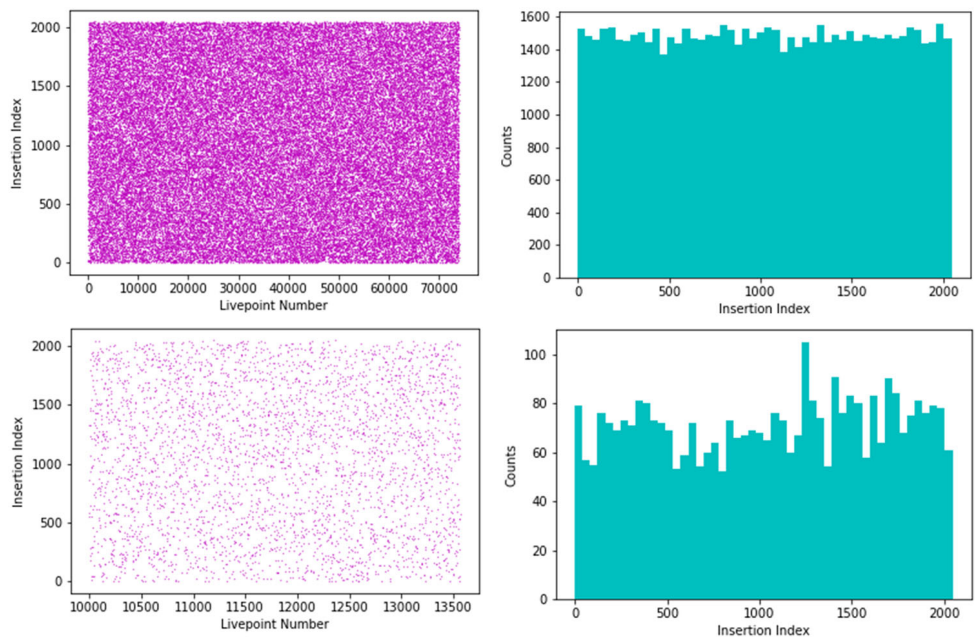

**Figure 2.** Top: example of the insertion order distribution from a single chain from GW150914 analyzed with IMRPhenomPv2. The flat distribution of insertion indices is typical for our dataset, and indicates that no major errors in prior sampling occurred in this chain. Bottom: a less uniform-looking example of a single chain from GW170823 analyzed with SEOBNRv4. The histogram shows one very frequent value, potentially indicating a plateau. However, in this case, the fluctuations are most likely due to the small-sample statistics of this shorter chain. The deviation from uniformity can be quantified using the KS-statistic and the significance of such a deviation can be assessed by calculating the corresponding $p$-value, which takes into account the sample size.

Having removed all chains with missing data and having transformed the insertion index to a continuous variable, we can now perform the final meta-test at the catalog level. **Figure 3** shows the results of the event-level $p$-value test, each conducted over all insertion index samples for each event, arranged in increasing order of KS $p$-value. The event names and their test results are listed in chronological order in Table 1. If the insertion order data were uniform, the $p$-values would also follow a uniform distribution, and the points in Figure 3 would fall along a straight line from 0 to 1. However, we still observe an excess of low $p$-values, particularly among events from O1 and O2. A final test comparing the meta $p$-values to the uniform distribution results in a KS $p$-value of $1.44 \times 10^{-3}$.

We also perform a meta-$p$-value test on the rolling test results. **Figure 4** shows the distribution of all rolling test $p$-values. The catalog-level meta-$p$-value, including all rolling tests and adjusted for the number, is 0.871, indicating no significant divergence from the uniform distribution and an overall healthy performance across all three runs.

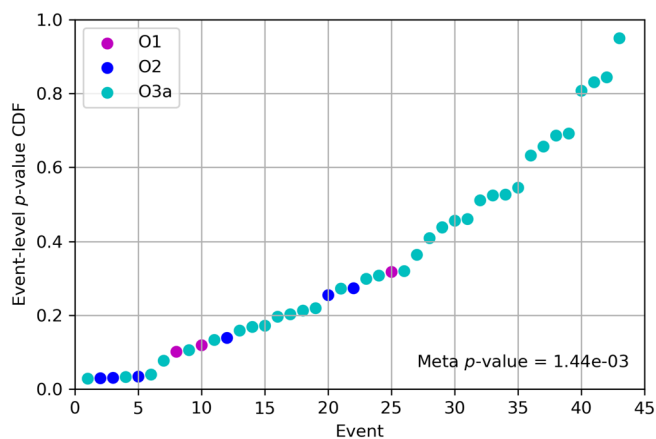

**Figure 3.** Cumulative plot of the event-level meta $p$-values associated with each event in the GWTC-1 and GWTC-2 dataset. Each point in this plot results from combining insertion orders of all chains from a single event and performing a KS test. If all insertion order data were completely uniform, the $p$-values would fall along a straight line from 0 to 1. The meta-$p$-value for this distribution is 0.00144.

## 6. Conclusion

We examined the hypothesis that the nested sampling analyses performed in GWTC-1 and GWTC-2 were well-behaved and unbiased from an insertion order statistics perspective. The event-level meta-$p$-values, shown in Figure 3 and listed in Table 1, follow a slightly non-uniform distribution at $p = 1.44 \times 10^{-3}$. In particular, although no individual pathological case has been identified,

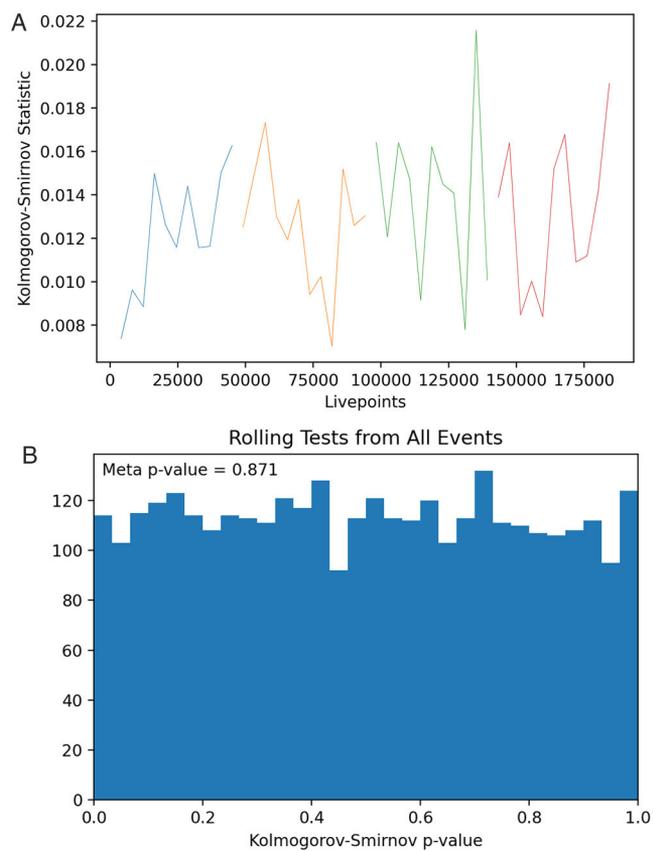

**Figure 4.** Top: typical example of a KS-statistic sequence obtained from rolling tests on data from GW190408, performed on chunks of $2 \times N_{\text{live}}$ points; data from each chain and log file is drawn in a different color. Bottom: histogram showing the distribution of the *p*-values of all rolling tests, conducted over all events and chains. Performing a catalog-level meta-test to compare this distribution to uniformity, we obtain a KS *p*-value of 0.871.

Figure 3 shows that there is a slight excess of lower *p*-values, providing weak evidence for sampling misbehavior in nested sampling. While the few O1 and O2 events tend to have lower meta *p*-values, the events from O3a cover the entire range from 0 to 1. This may indicate that, with updates to the sampling algorithms and more computational resources, the quality of nested sampling in LVC analysis is showing (weak) evidence of improvement over time. In particular, increasing the number of live points and the length of the MCMC chains used in prior sampling leads to more uniform insertion order statistics and, in consequence, to more reliable nested sampling output. The vast majority of the posterior samples is collected toward the final stages of the sampling, with dense exploration of the high-likelihood peak(s). So, there is little reason to believe that weak signs of misbehavior at random intervals on the chain could have a significant detrimental effect on the inference of the source parameters; however, they could impact evidence estimation.

The distribution of rolling test results pictured in Figure 4 is far more uniform than distribution of event-level meta test results, which is skewed toward lower *p*-values. The apparent contradiction between these results stems from the nature of the KS statistic. The KS test measures the supremum of distances from uniformity, so it measures whether any fault exists anywhere, not the average quality of sampling. While most individual segments are essentially uniform, it is unlikely that there would be no anomalies in multiple chains with tens of thousands of livepoints each. *p*-value adjustment for multiple tests reduces the impact of this effect, but does not eliminate it completely as long as the underlying distribution is nonuniform.

In combination, the rolling and meta tests indicate that sampling proceeds correctly in most small, local segments, but most events have at least one flaw in sampling somewhere. Moreover, we find that however tempting it is to make use of continuous tests of uniformity (such as the Kolmogorov–Smirnov or Anderson–Darling tests) without transforming discrete data, or to try and recover partial results from chains with missing data points, both of the above techniques lead to significant biases in our statistical results.

The location and *p*-value of the minimum rolling test can be used to identify and characterize these anomalous regions, and in practice, the combination of multiple chains and the supplementation of nested sampling with MCMC posteriors further suppress their effects. Taking all complete parallel chains into account, the overall results are consistent with unbiased nested sampling.

### Acknowledgements

The authors are grateful to John Veitch, Will Handley, Michael Williams, Ulrich Sperhake, Anthony Lasenby, Eugene Lim, and Roman Rafikov for their advice and feedback. The work of T.K. is supported by the Marshall Scholarship. The work of M.A. is supported by the Kavli Foundation. Computational resources for the data hosting and the data analysis calculations for this work were provided by the LIGO–Caltech Computing Cluster (National Science Foundation Grants PHY-0757058 and PHY-0823459), part of the LIGO Data Grid (LDG) and the International Gravitational-Wave Observatory Network (IGWN).

### Conflict of Interest

The authors declare no conflict of interest.

### Data Availability Statement

The data that support the findings of this study are available from the corresponding author upon reasonable request.

### Keywords

compact binaries, data analysis, gravitational waves, nested sampling

Received: June 12, 2022
Revised: July 11, 2022
Published online: